\documentclass[aps,prd]{revtex4}
\usepackage{amssymb}
\usepackage{graphicx}

\newcommand{\be}{\begin{equation}}
\newcommand{\ee}{\end{equation}}
\newcommand{\ba}{\begin{eqnarray}}
\newcommand{\ea}{\end{eqnarray}}
\newcommand{\ban}{\begin{eqnarray*}}
\newcommand{\ean}{\end{eqnarray*}}

\begin{document}

\title{\Large{What can we learn from electromagnetic plasmas about 
the quark-gluon plasma?}}

\author{Markus H. Thoma}

\affiliation{Max-Planck-Institut f\"ur extraterrestrische Physik,
P.O. Box 1312, 85741 Garching, Germany}



\begin{abstract}
Ultra-relativistic electromagnetic plasmas can be used for improving 
our understanding of the quark-gluon plasma. In the weakly coupled 
regime both plasmas can be described by transport theoretical and 
quantum field theoretical methods leading to similar results for the 
plasma properties (dielectric tensor, dispersion relations, plasma 
frequency, Debye screening, transport coefficients, damping and particle 
production rates). In particular, future experiments with ultra-relativistic 
electron-positron plasmas in ultra-strong laser fields might open the 
possibility to test these predictions, e.g. the existence of a new fermionic 
plasma wave (plasmino). In the strongly coupled regime electromagnetic 
plasmas such as complex plasmas can be used as models or at least analogies 
for the quark-gluon plasma possibly produced in relativistic heavy-ion 
experiments. For example, pair correlation functions can be used to 
investigate the equation of state and cross section enhancement for parton 
scattering can be explained.  
\end{abstract}

\maketitle


\section{Introduction}

At very high temperatures ($T> 150$ MeV) and densities ($\rho > 1$ GeV/fm$^3$)
it was predicted that QCD implies that there should be a phase transition from 
nuclear or hadronic matter to a system of deconfined quarks and gluons \cite{Collins}.
The early Universe should have been in this state for the first few microseconds
\cite{Collins}. Such a phase transition could also occur in the interior of neutron stars 
\cite{Ivanenko} or in relativistic nucleus-nucleus collisions leading to 
the so-called quark-gluon plasma (QGP) \cite{Shuryak}. The phase diagram 
is sketched in Fig.1.
The nature of this phase transition is similar to the Mott transition from an
insulator to a metal \cite{Mott}, where the insulator phase at low pressure corresponds
to the confined phase (nuclear matter) and the metallic phase at high pressure, e.g.
metallic hydrogen \cite{Wigner} predicted to exist in the interior of Jupiter \cite{Wildt},
to the deconfined phase (QGP). 

\begin{figure}
\begin{minipage}{6cm}
\includegraphics[width=3.0in]{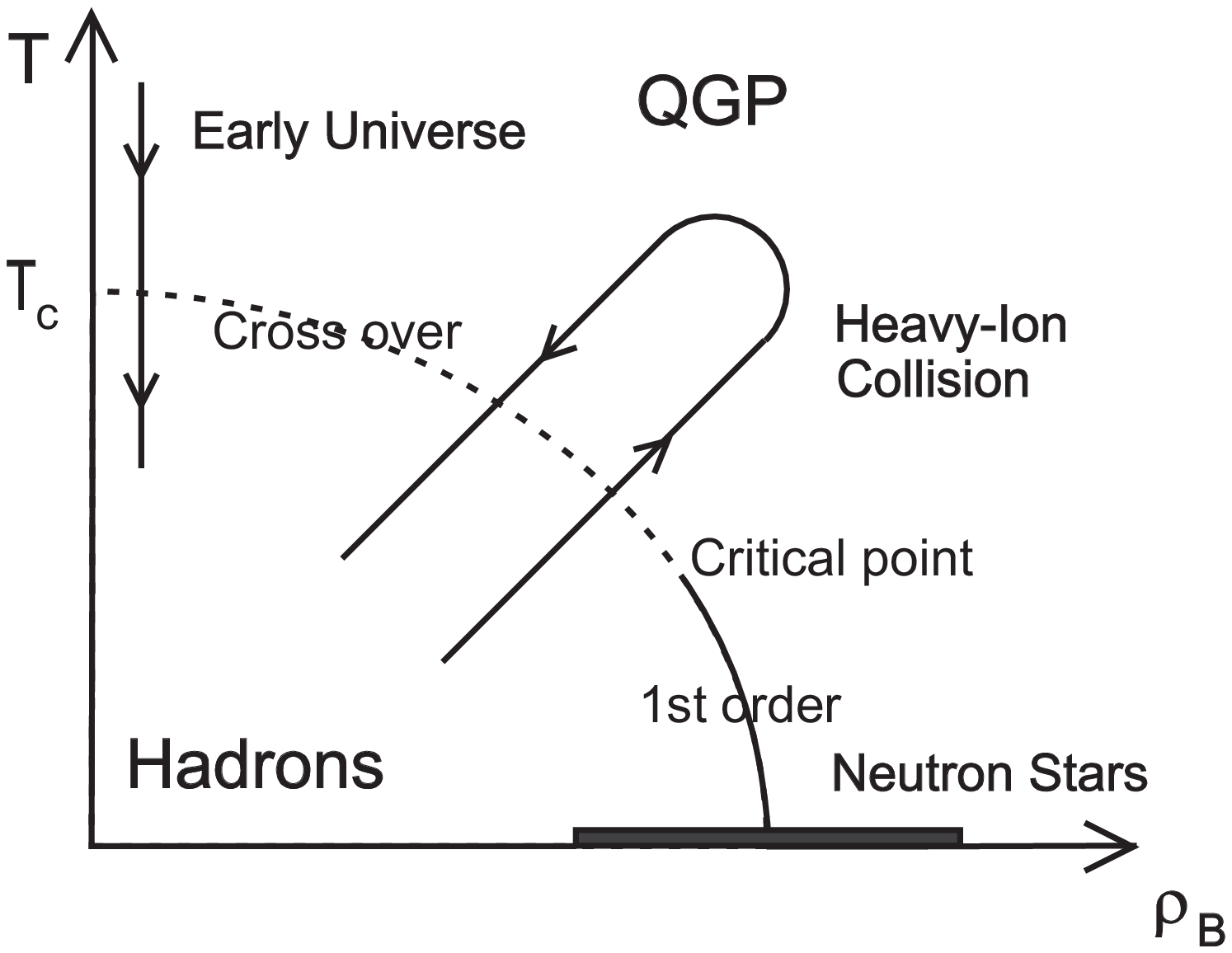}
\caption{Sketch of the QCD phase diagram}
\end{minipage}
\hfill
\begin{minipage}{9cm}
\includegraphics[width=3.5in]{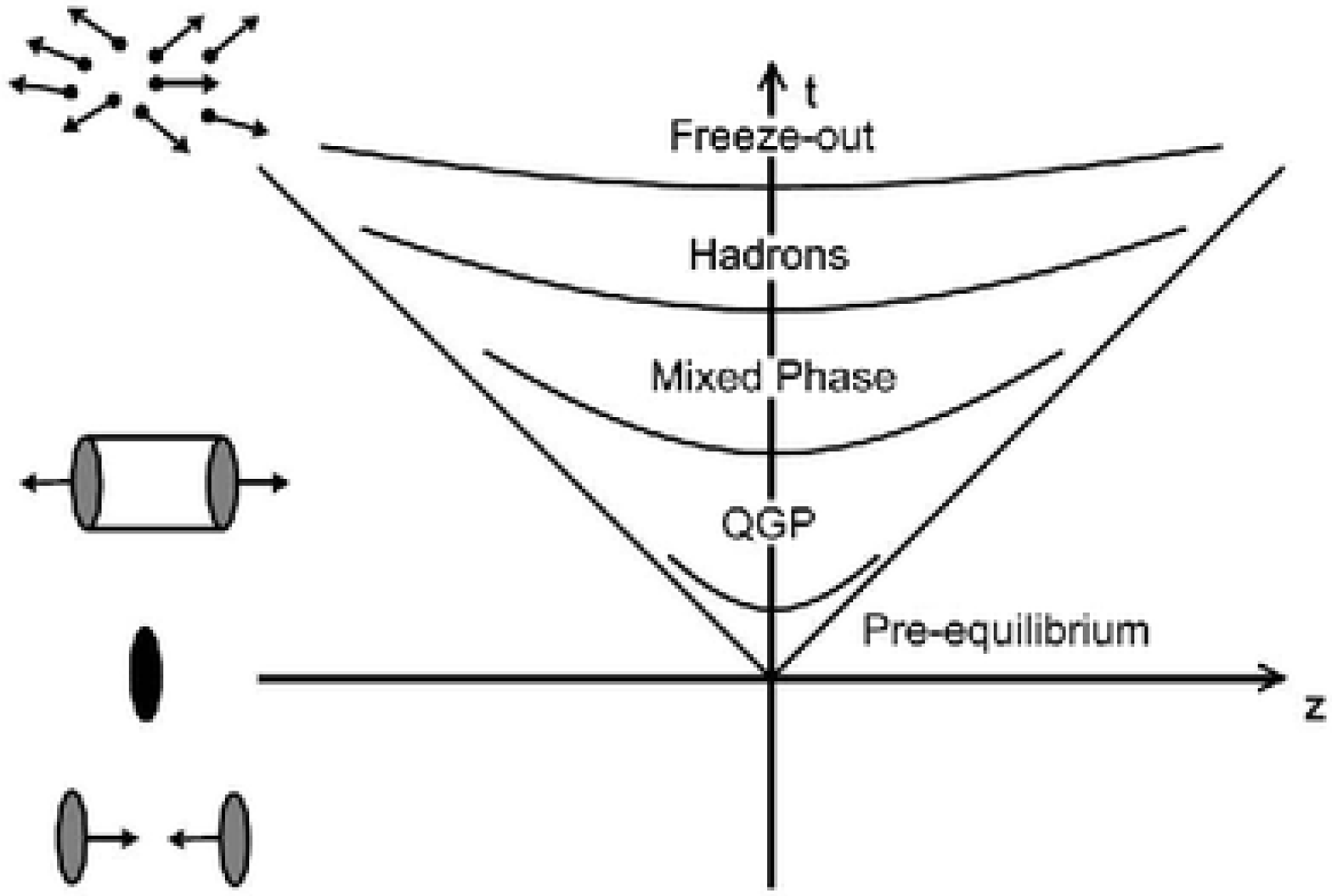}
\caption{Space-time evolution of the fireball in an ultrarelativistic heavy-ion collision}
\end{minipage}
\end{figure}

Estimates show that a QGP droplet
containing a few thousands quarks and gluons (partons) can be formed in the fireball of 
a heavy-ion collision which exists for a few fm/c. The space-time evolution of the fireball
looks like this (see Fig.2) \cite{Mueller}: after the collision of the two nuclei,
in which the nucleons dissolve into quarks and gluons, a pre-equilibrium phase
is present for about 1 fm/c which approaches equilibrium by secondary 
parton interactions. If the temperature and energy density  
are above the critical values, the fireball will be in the QGP phase, which
should be in thermal but maybe not in chemical equilibrium \cite{Biro}.
Due to the expansion the fireball will cool and the transition temperature
will be reached after a few fm/c followed by a mixed phase in case of
a first order transition and a hadronic phase. Finally the hadrons will cease
to interact if the system becomes dilute (freeze-out).

Since the QGP cannot be observed directly the big question is how to detect it.
Only by comparing theoretical predictions of signatures for the QGP formation
with experimental data, i.e. by circumstantial evidence, this goal can be achieved.
Hence a quantitative and detailed theoretical description of the QGP is required.
Basically there are three different methods: the first one, perturbative QCD, is
valid only for large parton momenta or extreme high temperatures  
for which the interaction between the partons becomes weak due to a property of QCD 
called asymptotic freedom. Perturbative QCD allows the calculation of static properties, 
e.g. equation of state, as well as dynamical quantities, e.g. parton scattering 
cross sections. The second method, lattice QCD is a truly non-perturbative method 
based on numerical simulations for solving the QCD equations on a discrete
4-dimensional space-time lattice. Unfortunately this technique can only be applied
to static quantities whereas most signatures of the QGP are dynamical like
particle production rates. The third method, which I would like to discuss
here in particular, is to apply and extend techniques from classical
electromagnetic plasmas to the QGP. For example, transport theory (Boltzmann
equation) or molecular dynamical simulations are widely used to describe properties
of plasmas. Also analogies with ideas and analytic models in plasma physics
can be useful. This I will explain in the next sections following partly the review
article \cite{Mrowczynski}. First I will start
with the weakly-coupled phase of the QGP, in which perturbative QCD can be used.
As we will see, the methods and results are very similar to a hot QED plasma
(electron-positron plasma) which can be investigated in the laboratory using
ultra-strong lasers in the near future and can serve therefore also as a test
model for the QGP.

\section{The Weakly-Coupled Quark-Gluon Plasma} 

Interactions between partons in the QGP lead to collective phenomena, 
such as Debye screening and plasma waves, or transport properties like viscosity.
At temperatures far above the critical temperature $T_c$ for the deconfinement
phase transition the effective temperature-dependent strong coupling constant, 
$\alpha_s=g^2/4\pi$, becomes small. At temperatures which can be reached
in heavy-ion experiments ($T<4 T_c$), $\alpha_s = 0.3 - 0.5$ is not really small
rendering the applicability of perturbation theory, which is an expansion in
the coupling constant, questionable. Perturbation theory in quantum field theory
is most conveniently done by using Feynman diagrams which can be via
Feynman rules directly translated into scattering amplitudes from which 
physical measurable quantities like cross sections, damping and production rates, 
and life times follow. If the interactions take place in the presence of a heat bath 
such as the QGP background, one has to consider QCD at finite temperature. For this 
purpose the Feynman rules have to be generalized to finite temperatures (and chemical
potential), which can be done either in the imaginary \cite{Matsubara} or
real \cite{Landsmann} time formalism. For an application of this method to the
QGP see for example Ref.\cite{Thoma}.

Alternatively transport theory can be used. For example, the dielectric tensor
of the QGP can be derived from combining the Vlasov and Maxwell equations. In an
isotropic QED plasma, in which there are two independent components, a longitudinal
and a transverse, one finds in this way \cite{Silin}
\ba
\epsilon_L(\omega ,k) & = & 1+\frac{3m_\gamma ^2}{k^2}\> 
\left[ 1-\frac{\omega}{2k} \> 
\ln \frac{\omega +k}{\omega -k}\right] \; ,
\nonumber \\[1mm]
\epsilon_T(\omega ,k) & = & 1-\frac{3m_\gamma ^2}{2k^2}\>
\left [1-\left (1-\frac{k^2}{\omega^2}\right )\> 
\frac{\omega}{2k}\> 
\ln \frac{\omega +k}{\omega -k} \right ]\;,
\label{e1}
\ea
where $\omega $ is the frequency, $k=|{\bf k}|$ the momentum, and $m_\gamma = eT/3$ the plasma
frequency. In the case of the QGP this result also holds if one replaces $m_\gamma $
by $m_g = \sqrt{(1+n_F/6)/3}\> gT$, where $n_F$ is the number of quark flavors
in the QGP. The Debye screening length following from the static limit ($\omega =0$)
of the longitudinal dielectric function is given by $\lambda_D=1/(\sqrt{3}\, m_\gamma)$.
For $\omega^2 < k^2$ the dielectric functions become negative corresponding to 
Landau damping.

The same result can be obtained from calculating the polarization tensor 
$\Pi_{\mu \nu}$ of Fig.3
in the high-temperature approximation \cite{Klimov,Weldon} and using the relation
\cite{Elze}
\ba
\epsilon_L(\omega ,k) & = & 1-\frac{\Pi_L (\omega ,k)}{k^2},
\nonumber \\[1mm]
\epsilon_T(\omega ,k) & = & 1-\frac{\Pi_T (\omega ,k)}{\omega^2},
\label{e2}
\ea
where $\Pi_L=\Pi_{00}$ and $\Pi_T=\sum_{ij} (\delta_{ij}-k_ik_j/k^2)\Pi_{ij}/2$
with $i$ and $j$ representing the space indices. 

\begin{figure}
\begin{minipage}{6cm}
\includegraphics[width=2.5in]{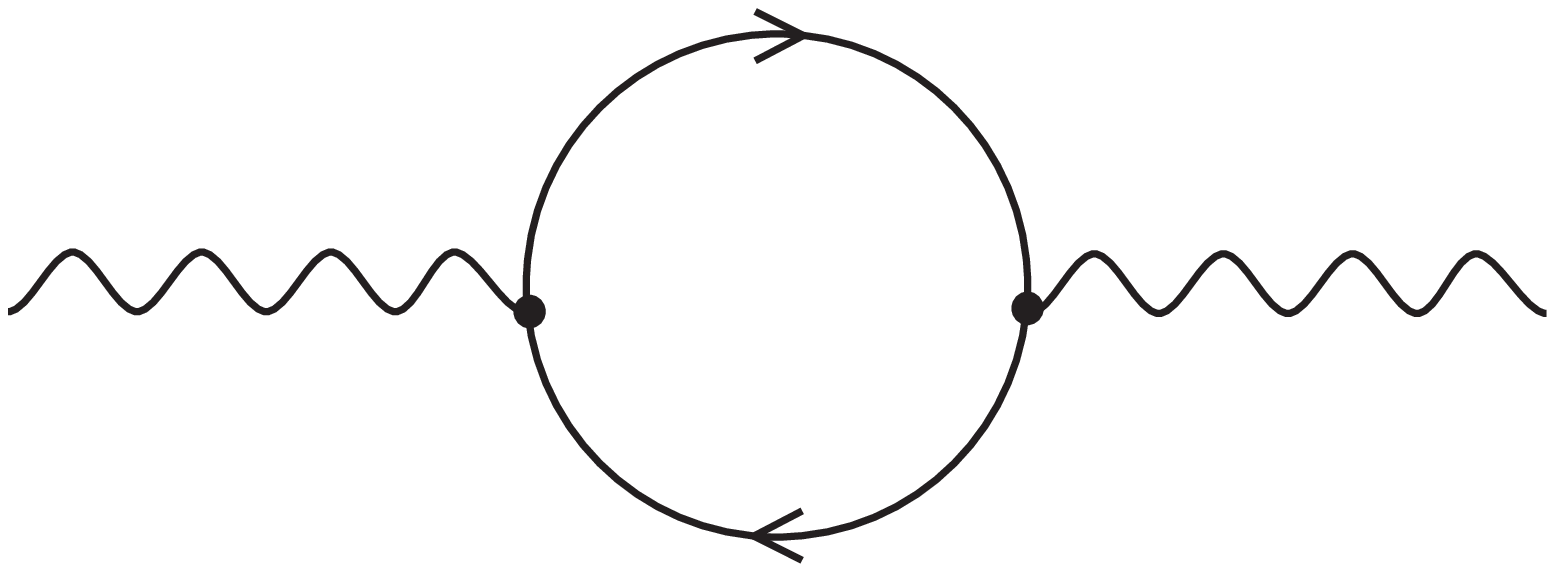}
\caption{Polarization tensor in lowest order perturbation theory}
\end{minipage}
\hfill
\begin{minipage}{10cm}
\includegraphics[width=4.0in]{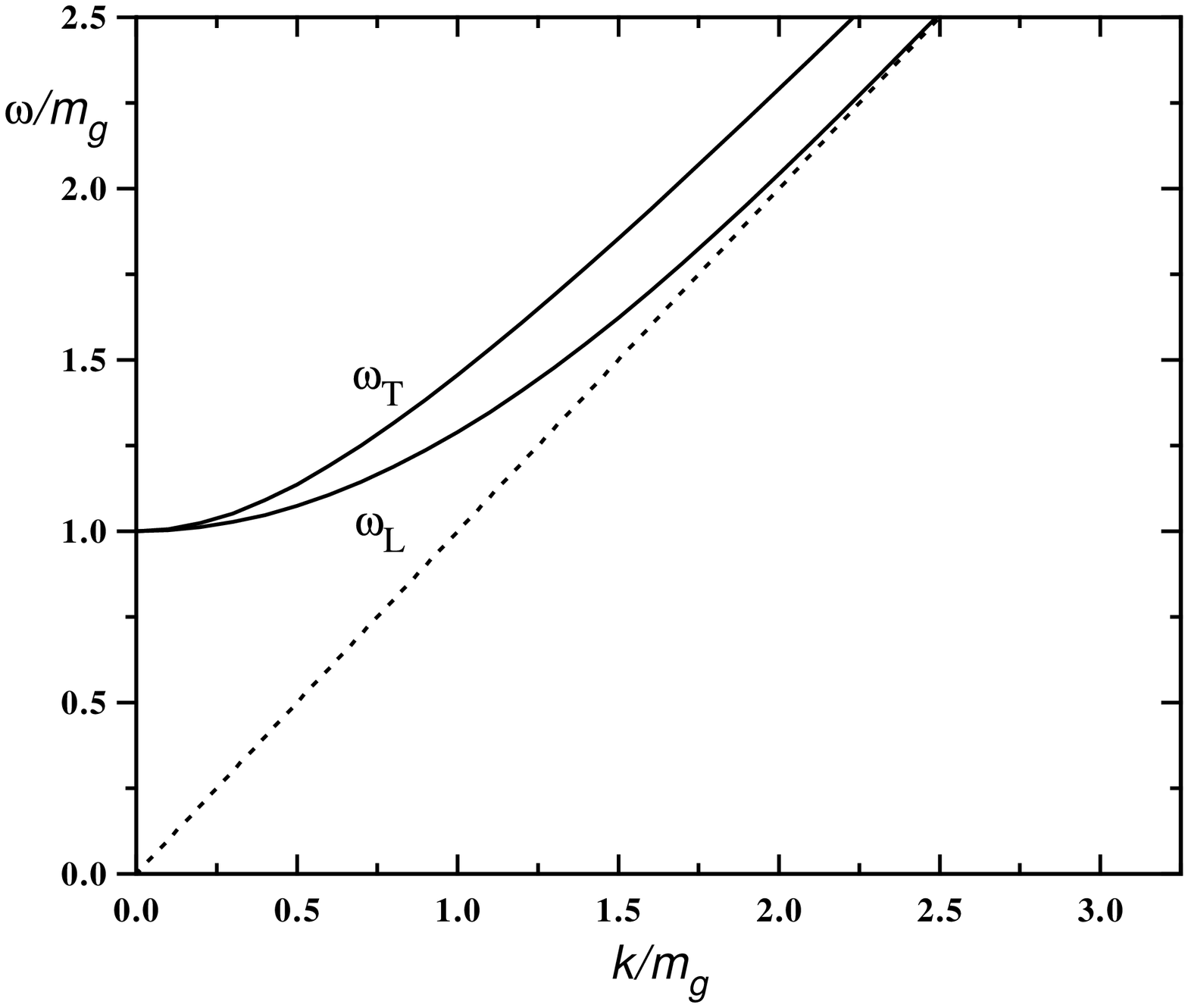}
\caption{Dispersion relations of photons in a QED plasma or gluons in a QGP}
\end{minipage}
\end{figure}

Using the Maxwell equations the dispersion relations $\omega_{L,T}(k)$ of plasma waves,
describing the propagation of electromagnetic, i.e. photons, (or chromoelectromagnetic,
i.e. gluons, in the case of a QGP),
waves in the plasma, can be found from
\ba
\epsilon_L(\omega ,k) & = & 0,
\nonumber \\[1mm]
\epsilon_T(\omega ,k) & = & \frac{k^2}{\omega^2}.
\label{e3}
\ea 
The dispersion relations are shown in Fig.4, where the branch $\omega_L$ is called
a plasmon, representing a longitudinal electromagnetic wave in the medium which does not exist in 
vacuum. In a relativistic plasma the transverse waves associated with the magnetic
interaction are as important as the longitudinal ones (plasmons), whereas in non-relativistic
plasmas only the plasmons are of significance. If one wants to go beyond the high-temperature
approximation, one cannot use classical transport theory but has to consider higher order diagrams
for the polarization tensor. 

Further interesting quantities can be derived from the dielectric tensor. For example,
the wake potential of a moving charge $Q$ with velocity ${\bf v}$, such as a heavy quark 
propagating through the QGP with a large transverse momentum from initial hard collisions,  
is given by \cite{MTC}
\be 
\phi({\bf r}, t, {\bf v})=\frac{Q}{2\pi^2}\> \int d^3k \frac{e^{-i{\bf k}\cdot ({\bf r}-{\bf v}t)}}
{k^2 \epsilon_L(\omega={{\bf v}\cdot {\bf k}, k})}.
\label{e4}
\ee
Wakes created by fast quarks in a QGP lead to an attraction and the possible
formation of diquarks \cite{Chakraborty}. Wakes and Mach cones have been
observed also in complex or dusty plasmas \cite{Samsonov}. A closely related quantity 
is the energy loss of an energetic quark or gluon in the QGP. Partons with a large 
transverse momentum $p_T$ are created from initial hard collisions between the partons
in the nuclei. Those quarks and gluons propagate through the fireball losing
energy. The energy loss coming from the back reaction of the induced electric field
of a fast moving charge in a plasma can be expressed by the dielectric functions
of the plasma \cite{Ichimaru}. In the case of a high energy quark with velocity $v$
in the QGP this energy loss is given by \cite{Thoma1}
\be 
\frac{dE}{dx}=-\frac{4\alpha_s}{3\pi v^2}\> \int \frac{dk}{k} \int_{-vk}^{vk} d\omega 
\omega \> \left [Im \frac{1}{\epsilon_L(\omega , k)} + (v^2 k^2 -\omega^2) Im \frac{1}
{\omega^2 \epsilon_T(\omega , k) -k^2} \right ].
\label{e5}
\ee
In addition there are contributions to the energy loss from elastic parton scattering and
from bremsstrahlung (radiative energy loss) \cite{Baier}. The combination of the 
energy loss by the induced electric field, corresponding to long-range interactions within 
the plasma, together with individual elastic scatterings is called collisional
energy loss. A consistent perturbative treatment of the collisional energy loss based 
on the hard thermal loop resummation technique \cite{Braaten1} was presented in
Ref.\cite{Braaten2}. Later on it was argued that the radiative energy loss dominates 
for relativistic partons in the QGP (see e.g. Ref.\cite{Baier}). However, it could be shown
\cite{Mustafa1,Wicks} that for the quenching of high-$p_T$ hadron spectra resulting from the
energy loss the collisional and radiative energy loss contribute equally. Indeed, the
quenching of hadron spectra observed at RHIC is significantly larger than predicted solely 
from the radiative energy loss, which was accepted as a signature for the formation of
a QGP phase in relativistic heavy-ion collisions at RHIC as the energy loss in hadronic matter
is assumed to be smaller \cite{Wicks}.

The dielectric functions (\ref{e1}) are derived from the Vlasov equation, i.e. for
the case of a collisionless plasma. Of course, in most cases collisions play an important
role in a plasma, e.g. for equilibration and transport properties, and cannot be neglected.
Collisions can be considered by using the Boltzmann equation (Vlasov equation plus collision 
term). There are several approximations to this equation, of which the relaxation time 
approximation replacing the collision integral by a constant collision rate $\nu$ is the
simplest one. In particular the formalism proposed in Ref.\cite{Bhatnagar} has been applied
to low-temperature plasmas successfully (for complex plasmas see e.g. Ref.\cite{Khrapak}). 
In this case an analytic
expression for the dielectric functions containing the collision rate can be derived
for ultrarelativistic plasmas \cite{Carrington}. For example, the longitudinal
dielectric function now reads
\be
\epsilon_L(\omega ,k) = 1+\frac{3m_\gamma ^2}{k^2}\> 
\left[ 1-\frac{\omega +i\nu}{2k} \> 
\ln \frac{\omega +i\nu +k}{\omega +i\nu -k}\right]\left[ 1-\frac{i\nu}{2k} \> 
\ln \frac{\omega +i\nu +k}{\omega +i\nu -k}\right]^{-1}.
\label{e6}
\ee
The plasmon dispersion relation now cuts the light cone ($\omega =k$) at a finite
value of $k$ and the plasma frequency is given by $\omega_L(k=0)<m_\gamma$. In the
case of a QGP $m_\gamma$ is replaced again by $m_g$. 

In heavy-ion collisions the fireball expands dominantly in longitudinal (beam)
direction. Hence the system is not isotropic and more components of the dielectric tensor
have to be considered \cite{Mrowczynski1}. In anisotropically expanding plasmas instabilities
such as the Weibel instability show up. It has been shown that these instabilities can drive
a QGP rapidly to equilibrium even in the weakly coupled case \cite{Mrowczynski2}. However, 
using the relaxation time approach as in (\ref{e6}) it was argued that these instabilities
are suppressed to some extent by the presence of collisions in the plasma \cite{Schenke}.   

\begin{figure}
\begin{minipage}{6cm}
\includegraphics[width=2.5in]{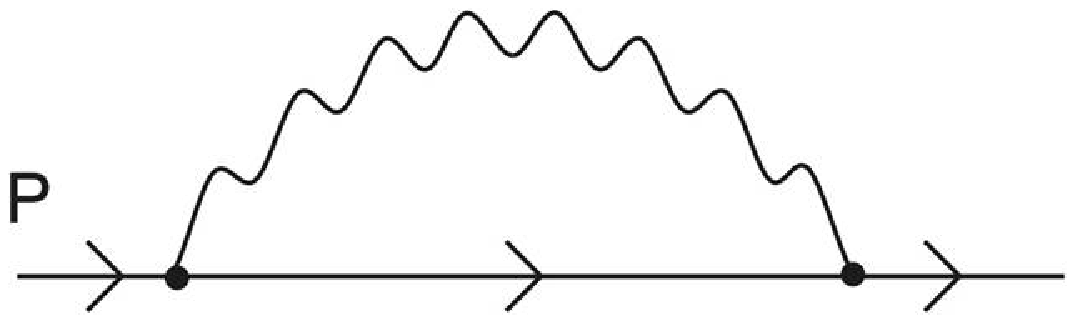}
\caption{Fermion self-energy in lowest order perturbation theory}
\end{minipage}
\hfill
\begin{minipage}{10cm}
\includegraphics[width=4.0in]{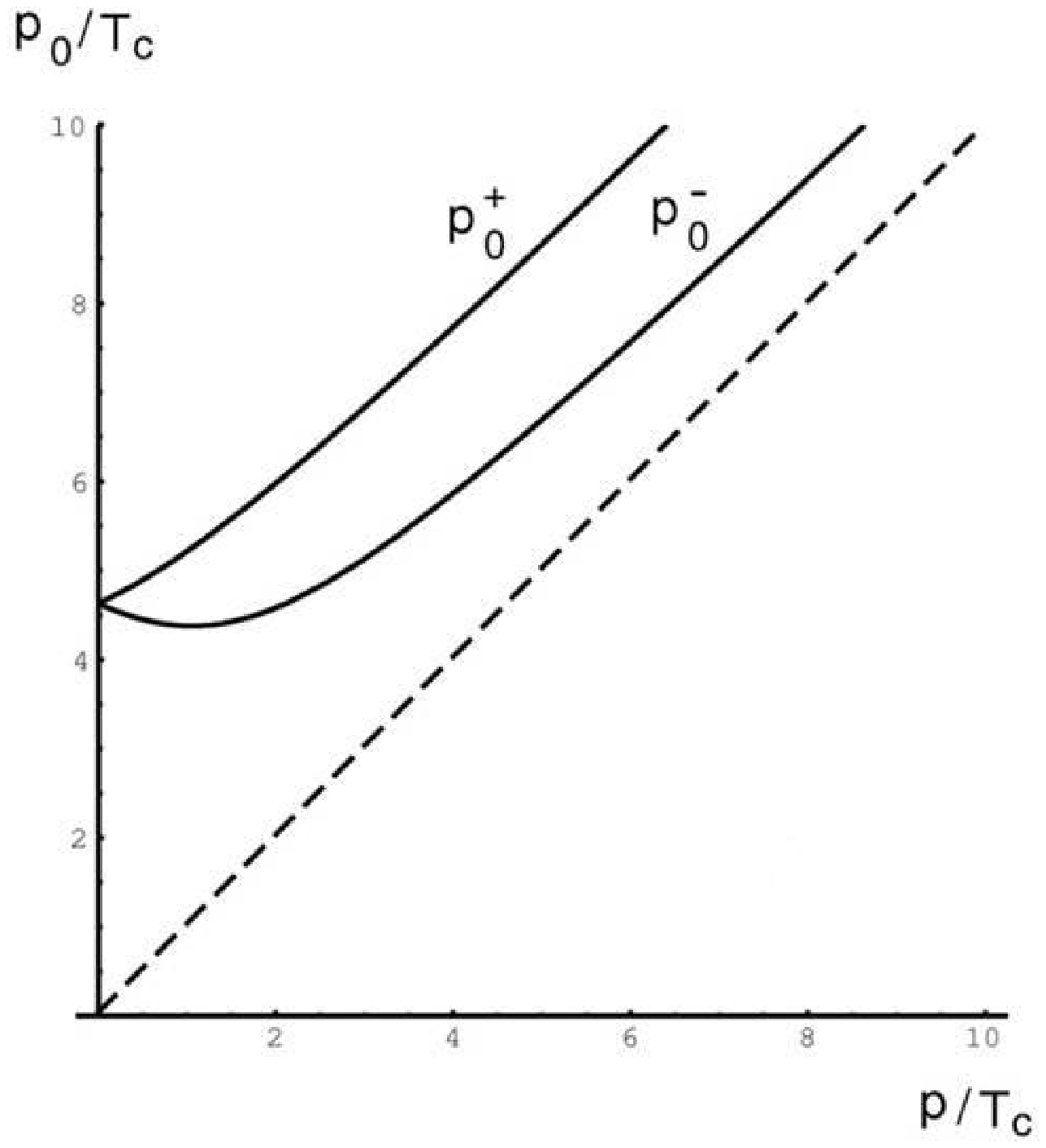}
\caption{Dispersion relations of electrons in a QED plasma or quarks in a QGP}
\end{minipage}
\end{figure}
 
Ultrarelativistic plasmas offer the exciting possibility to observe a new kind of collective 
modes besides electromagnetic ones, namely fermionic modes. They cannot be derived from
the dielectric function but from the electron or quark self-energy. In lowest order perturbation
theory the fermion self-energy is given by the diagram in Fig.5. The pole of the effective
fermion propagator following from a resummation of this self-energy describes the dispersion relation
of collective fermion modes in the plasma. As in the case of electromagnetic (photon or gluon) 
waves there are two branches, one with a positive ratio of helicity to chirality and one
with a negative, which is absent in vacuum, called plasmino \cite{Braaten3}. The plasmino branch 
shows a minimum at a finite value of the momentum $p$ as sketched in Fig.6. A minimum in the 
dispersion relation leads to van Hove singularities which could show up in sharp peaks
in the dilepton production rate in the QGP, serving as a possible signature for the QGP
formation \cite{Braaten3}. It was argued that this minimum in the plasmino branch
is a general feature of 
ultrarelativistic plasmas independent of the approximation, e.g. perturbation theory
\cite{Peshier}. The observation of collective electron modes in an ultrarelativistic
electron-positron plasma, produced by strong laser fields in the laboratory \cite{Shen}, 
would be an exciting discovery also serving as a test for the QGP \cite{Thoma2}.  

As a side remark, I would like to mention that the fermion self-energy can also be calculated
perturbatively at high quark densities by introducing a finite chemical potential $\mu$. In this way, 
for example, a quasi-particle mass for quarks in quark matter can be derived which describes the
equation of state of a quasi-particle Fermi gas. The mass-radius relation for quark stars has been
computed using this equation of state \cite{Schertler}. However, so far no indications for 
quark stars have been found (see e.g. \cite{Thoma3}).  

There are a number of other interesting quantities which can be calculated by perturbation theory
going beyond the classical high-temperature approximation. For this purpose, one has to adopt
the hard thermal loop resummation technique in order to obtain consistent, i.e. complete to 
leading order, infrared finite, and gauge independent, results \cite{Braaten1}. Important examples are
damping rates, transport rates, mean free paths, collision times, and transport coefficients 
(viscosity) of electrons, quarks, photons, and gluons \cite{Thoma,Thoma2}.
Of course, these results hold only in the limit of extremely high temperatures - even at the
Planck scale the QCD coupling constant $g$ is of the order of 1/2. However, they might provide 
qualitative insight into important aspects of the physics of the QGP such as the role of 
collective effects. In the case of an electron-positron plasma, however, 
these calculations are reliable, serving as predictions for such a QED plasma which might be 
produced in strong laser fields soon \cite{Thoma2}.
 	 
\section{The Strongly-Coupled Quark-Gluon Plasma}

The essential parameter distinguishing between weakly-coupled and strongly coupled 
non-relativistic plasmas is the Coulomb coupling parameter defined by the ratio
of the interaction energy (Coulomb energy) between the plasma particles and 
their thermal energy \cite{Ichimaru2}
\be
\Gamma =\frac{Q^2}{d k_BT},
\label{e7}
\ee
where $Q$ is the charge of the particles, $d$ the interparticle distance, and $T$ 
the plasma temperature. For strongly-coupled plasmas this parameter is of the
order of one or larger. For example, for the ion component in a white dwarf
$\Gamma$ can be between about 5 and 500. In complex or dusty plasmas, which contain
micron size particles, e.g. dust grains, the microparticle component 
is highly charged (several thousand elementary charges) due to electron collection
and interacts via a Yukawa potential
leading to a $\Gamma$ between 1 and $10^5$ depending on the particle size and
the plasma parameters \cite{Fortov}. A simple, extensively studied
model for a strongly coupled plasma is the one-component plasma, in which
particles of the same charge in a neutralizing background interact via a
Coulomb potential \cite{Ichimaru2}. For $\Gamma >1$ the plasma shows a liquid-like 
behavior (see below), and for $\Gamma >172$ a crystalline structure. Such a plasma 
crystal has been observed in complex plasmas in the laboratory \cite{Thomas}.
An improvement of the one-component model is the Yukawa
system taking into account the Debye screening of the particle charge.

In the case of a QGP the interaction parameter was estimated to be \cite{Markus}
\be
\Gamma =2 \frac{C\alpha_s}{d k_BT},
\label{e8}
\ee
where $C=4/3$ is the Casimir invariant in the case of quarks and $C=3$ in the case
of gluons. The pre-factor of 2 has been added to consider the fact that  
in relativistic plasmas the magnetic interaction is as important as the electric.
Employing realistic values for RHIC energies, e.g. $T=200$ MeV, $\alpha_s=0.3 - 0.5$,
and $d=0.5$ fm, we find $\Gamma = 1.5 - 6$. Here it should be noted that only the
Coulomb potential corresponding to a one-gluon exchange was assumed. Higher order
and non-perturbative effects can increase the value of $\Gamma$ significantly.
Anyway this estimate indicates that the QGP in ultrarelativistic heavy-ion collisions
is a strongly-coupled plasma probably in the liquid phase. This means there could be a 
phase transition to a QGP gas at higher temperatures where $\Gamma $ will be smaller \cite{Markus1}.
However, only in the simultaneous presence of attractive and repulsive interactions,
such as a Lennard-Jones Potential, this phase transition is of first order with a
critical end point. Otherwise, the system is always in the supercritical phase allowing
no determination of a sharp borderline between the liquid and the gaseous behavior.

The theoretical description of the strongly-coupled QGP is difficult because perturbative
QCD is not applicable and lattice QCD is restricted to static quantities 
and its accuracy at finite temperature - not to speak of finite chemical potential - 
not yet satisfactory in many cases. Therefore electromagnetic strongly-coupled plasmas, 
which can be investigated much easier, and the methods, e.g. molecular dynamics 
\cite{Gelman,Hartmann}, for describing them are considered to 
improve our - at least qualitative - understanding of the QGP by analogy. For example,
ultracold quantum gases exhibit a similar behavior in the flow pattern observed at
RHIC \cite{Shuryak1}. The elliptic flow investigated in these heavy-ion collisions 
can be described very well
by ideal hydrodynamics, indicating the presence of an almost ideal QGP liquid \cite{McLerran}.
Also high-density plasmas produced by shooting heavy-ion beams onto solid state targets
are another example of strongly-coupled plasmas in the laboratory \cite{Dewald}. A model system
for the QGP which can be produced for a wide range of values of $\Gamma$ and investigated on 
the microscopic and dynamical level in real time by direct optical observation is the complex 
plasma, discussed already above. It has the further advantage that the strong-coupling is due 
to the large coupling (high charge of the microparticles) as in the case of the QGP and
not because of high density or low temperature. 
   
An important tool for investigating fluids, such as complex plasmas in the liquid phase, 
on the microscopic level are the pair
correlation function and its Fourier transform the static structure function 
\cite{Hansen}. The qualitative behavior of the latter for the gas and liquid phase
is shown in Fig.7. Using the hard thermal loop approximation, valid in the weak
coupling limit, a gas-like behavior of an interacting quark system is found \cite{Thoma5},
\be
S(p)=\frac{2n_FT^3}{n}\> \frac{p^2}{p^2+m_D^2},
\label{e9} 
\ee
where $n$ is the particle density, $n_F$ the number of quark flavors, and
$m_D=1/\lambda_D$ the inverse Debye screening length. It would be interesting to compute
the static structure function non-perturbatively for realistic situations, in particular 
by using lattice QCD, to see whether an oscillatory behavior indicating a QGP liquid
will be found. 

\begin{figure}
\includegraphics[width=3.5in]{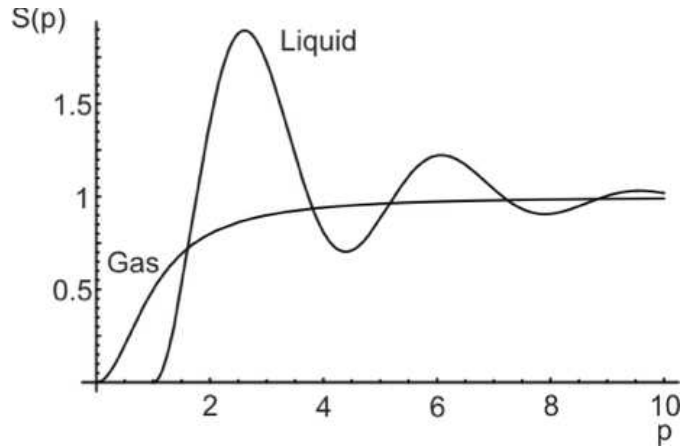}
\caption{Qualitative behavior of the static structure function for the liquid and gaseous
phase}
\end{figure}

Another example of a lesson that can be learned from strongly-coupled electromagnetic
plasmas is the cross section enhancement of the particle scattering due to 
non-linear effects. It has been argued that RHIC data imply a cross section
enhancement in the elastic parton scattering by an order of magnitude compared
to perturbative results \cite{Molnar}. Indeed in strongly-coupled plasmas,
such as complex plasmas \cite{Khrapak2}, such a cross section enhancement
takes place because the interaction range in these plasmas is larger than the
Debye screening length which therefore cannot be used as an infrared
cutoff in the calculation of the absolute cross section. An estimate of this
effect in the QGP case gives an parton cross section enhancement by a factor 2 - 9
implying a small mean free path, a small viscosity and a fast thermalization in accordance
with RHIC experiments \cite{Markus}. Furthermore, such a cross section enhancement would lead
to a further increase of the collisional energy loss, given approximately
by the energy transfer per collision divided by the mean free path. The radiative
energy loss, on the other hand, could be suppressed due to an enhancement of the
Landau-Pomeranchuk-Migdal effect, which describes a suppression of the photon
or gluon radiation if the time between two scatterings is too small to allow the 
emission of the photon or gluon \cite{Markus1}.

As a last example for an interesting comparison between the QGP and strongly-coupled
electromagnetic plasmas let me mention the prediction of a lower limit for the ratio
of viscosity to entropy density from string theory (AdS/CFT) \cite{Kovtun},
\be
\frac{\eta}{s} \leq \frac{\hbar}{4\pi k_B},
\label{e10}
\ee
which is
widely discussed in the QGP community because the QGP seems to come close to this limit
\cite{Romatschke}. In general, strongly interacting systems show a small viscosity,
e.g. the one-component plasma has a minimum in the viscosity at $\Gamma =21$ \cite{Saigo}.
The minimum of the ratio of viscosity to entropy density in the one-component plasma at
$\Gamma = 12$ is about 5 times above the string theory limit \cite{Thoma6} - water under
normal conditions exceeds this limit by about a factor of 400 - similar to 
predictions for the QGP.

\section{Conclusions}

Transport theoretical methods (Vlasov and Boltzmann equation) widely used for 
non-relativistic electromagnetic plasmas and perturbative field theory at finite temperature 
(and chemical potential) can be used
for describing the weakly coupled phase of the QGP, e.g. collective and transport
properties, and high-energy phenomena (jets, hard photons) in the QGP. Also properties 
of a relativistic electron-positron plasma produced in ultrastrong laser fields
or supernovae can be treated in this way. The comparison between these two systems
might be helpful to learn about the role of the strong coupling effects in the QGP.
Field theoretic predictions of a new phenomenon not known from non-relativistic plasmas, 
namely the existence of collective fermion modes (plasminos), might open exciting
investigations of relativistic electron-positron plasmas in the laboratory.

Properties of a strongly-coupled QGP such as a liquid phase, cross section enhancement, and
small viscosity can be studied in analogy to strongly-coupled electromagnetic plasmas.
In particular complex plasmas, which can be easily produced with a highly tunable
interaction strength and directly investigated on the microscopic and dynamical level,
showing a large variety of interesting features such as solid and liquid phases, 
might be useful in this respect.  

\bigskip



\begin{thebibliography}{1}
\bibitem{Collins} J.C. Colins and M.J. Perry, Phys. Rev. Lett. 34 (1975) 1353.
\bibitem{Ivanenko} D.D. Ivanenko and D.F. Kurdgelaidze, Sov. Phys. J. 13 (1970) 1015.
\bibitem{Shuryak} E.V. Shuryak, Phys. Lett. 78B (1978) 150.
\bibitem{Mott} N.F. Mott, Rev. Mod. Phys. 40 (1968) 677.
\bibitem{Wigner} E. Wigner and H.B. Huntington, J. Chem. Phys. 3 (1935) 764.
\bibitem{Wildt} R. Wildt, Ap. J. 87 (1938) 508.
\bibitem{Mueller} B. M\"uller, {\it The Physics of the Quark-Gluon Plasma}, 
Lecture Notes in Physics 225 (Springer, Berlin, 1985).
\bibitem{Biro} T.S. Biro, E. van Doorn, B. M\"uller, M.H. Thoma, and X.N. Wang,
Phys. Rev. C 48 (1993) 1275.
\bibitem{Mrowczynski} S. Mrowczynski and M.H. Thoma, Annu. Rev. Nucl. Part. Sci.
57 (207) 61.
\bibitem{Matsubara} T. Matusbara, Progr. Theor. Phys. 14 (1955) 351.
\bibitem{Landsmann} N.P. Landsmann and C.G. van Weert, Phys. Rep. 145 (1987) 141.
\bibitem{Thoma} M.H. Thoma, in: {\it Quark Gluon Plasma 2}, ed. R.C. Hwa (World Scientific,
Singapore, 1995), p.51, {\it hep-ph/9503400}. 
\bibitem{Silin} V.P. Silin, Sov. Phys. JETP. 11 (1960) 1136.
\bibitem{Klimov} V.V. Klimov, Sov. Phys. JETP 55 (1982) 199.
\bibitem{Weldon} H.A. Weldon, Phys. Rev. D 26 (1982) 1394.
\bibitem{Elze} H.T. Elze and U. Heinz, Phys. Rep. 138 (1989) 81.
\bibitem{MTC} M.G. Mustafa, M.H. Thoma, and P. Chakraborty, Phys. Rev. C 71 (2005) 017901.
\bibitem{Chakraborty} P. Chakraborty, M.G. Mustafa, and M.H. Thoma, Phys. Rev. D 74 (2006) 094002.
\bibitem{Samsonov} D. Samsonov, J. Goree, Z.W. Ma, A. Bhattacharjee, H.M. Thomas, and G.E. Morfill,
Phys. Rev. Lett. 83 (1999) 3649.
\bibitem{Ichimaru} S. Ichimaru, {\it Basic Principles of Plasma Physics} (Benjamin, 1973, Reading).
\bibitem{Thoma1} M.H. Thoma and M. Gyulassy, Nucl. Phys. B 351 (1991) 491.
\bibitem{Baier} R. Baier, D. Schiff, and B.G. Zakharov, Annu. Rev. Nucl. Part. Sci. 50 (2000) 37.
\bibitem{Braaten1} E. Braaten and R.D. Pisarski, Nucl. Phys. 337 (1990) 569.
\bibitem{Braaten2} E. Braaten and M.H. Thoma, Phys. Rev. D 44 (1991) 1298, {\it ibid.} R2625.
\bibitem{Mustafa1} M.G. Mustafa and M.H. Thoma, Acta Phys. Hung. A22 (2005) 93; M. Mustafa, Phys.
Rev. C 72 (2005) 014905.
\bibitem{Wicks} S. Wicks, W. Horowitz, M. Djordjevic, and M. Gyulassy, Nucl. Phys. A784 (2007) 426. 
\bibitem{Bhatnagar} P.L. Bhatnagar, E.P. Gross, and M. Krook, Phys. Rev. 94 (1954) 511.
\bibitem{Khrapak} S.A. Khrapak {\it et al.}, Phys. Rev. E 72 (2005) 016406.
\bibitem{Carrington} M.E. Carrington, T. Fugleberg, D. Pickering, and M.H. Thoma, Can. J. Phys. 82
(2004) 671.
\bibitem{Mrowczynski1} S. Mrowzcynski and M.H. Thoma, Phys. Rev. D 62 (2000) 036011.
\bibitem{Mrowczynski2} J. Randrup and S. Mrowczynski, Phys. Rev. C 68 (2003) 034909.
\bibitem{Schenke} B. Schenke, M Strickland, C. Greiner, and M.H. Thoma, Phys. Rev. D 73 (2006) 125004.
\bibitem{Braaten3} E. Braaten, R.D. Pisarski, and T.C. Yuan, Phys. Rev. Lett. 64 (1990) 2242.
\bibitem{Peshier} A. Peshier and M.H. Thoma, Phys. Rev. Lett. 84 (2000) 841.
\bibitem{Shen} B. Shen and J. Meyer-ter-Vehn, Phys. Rev. E 65 (2001) 016405.
\bibitem{Thoma2} M.H. Thoma, {\it arXiv:0801.0956}, to be published in Rev. Mod. Phys.
\bibitem{Schertler} K. Schertler, C. Greiner, J. Schaffner-Bielich, and M.H. Thoma,
Nucl. Phys. A677 (2000) 463.
\bibitem{Thoma3} M.H. Thoma, J. Tr\"umper, and V. Burwitz, J.Phys. G30 (2004) S471.
\bibitem{Ichimaru2} S. Ichimaru, Rev. Mod. Phys. 54 (1982) 1017.
\bibitem{Fortov} V.E. Fortov, A.V. Ivlev, S.A. Khrapak, A.G. Khrapak, and G.E. Morfill,
Phys. Rep. 421 (2005) 1.
\bibitem{Thomas} H.M. Thomas, G.E. Morfill, V. Demmel, J. Goree, B. Feuerbacher, and 
D. M\"ohlmann, Phys. Rev. Lett. 73 (1994) 652.
\bibitem{Markus} M.H. Thoma, J. Phys. G 31 (2005) L7 and Erratum {\it ibid.} 539.
\bibitem{Markus1} M.H. Thoma, Nucl. Phys. A774 (2006) 307.
\bibitem{Gelman} B.A. Gelman, E.V. Shuryak, and I. Zahed, Phys. Rev. C 74 (2006) 044908.
\bibitem{Hartmann} P. Hartmann, Z. Donko, P. Levai, and G.J. Kalman, Nucl. Phys. A774 (2006) 881.
\bibitem{Shuryak1} E. Shuryak, {\it arXiv:0807.3033}.
\bibitem{McLerran} M. Gyulassy and L. McLerran, Nucl. Phys. A750 (250) 30.  
\bibitem{Dewald} E. Dewald {\it et al.}, IEEE Trans. Plasma Sc. 31 (2003) 221.
\bibitem{Hansen} J.-P. Hansen and I.R. McDonald, {\it Theory of Simple Liquids}
(2nd edition, Academic Press, London, 1986).
\bibitem{Thoma5} M.H. Thoma, Phys. Rev. D 72 (2005) 094030.
\bibitem{Molnar} D. Molnar and M. Gyulassy, Nucl. Phys. A697 (2002) 495.
\bibitem{Khrapak2} S.A. Khrapak, A.V. Ivlev, G.E. Morfill, and H.M. Thomas,
Phys. Rev. E 66 (2002) 046414.
\bibitem{Kovtun} P.K. Kovtun, D.T. Son, and A.O. Starinets, Phys. Rev. Lett. 94 (2005) 111601.
\bibitem{Romatschke} P. Romatschke and U. Romatschke, Phys. Rev. Lett. 99 (2007) 172301. 
\bibitem{Saigo} T. Saigo and S. Hamaguchi, Phys. Plasmas 9 (2002) 1210.
\bibitem{Thoma6} M.H. Thoma and G.E. Morfill, Eur. Phys. Lett. 82 (208) 65001.

\end{thebibliography}
\end{document}